*On the origin of surface imposed anisotropic growth of salicylic and acetylsalicylic acids crystals during droplet evaporation*

**Maciej Przybyłek, Piotr Cysewski, Maciej Pawelec, Dorota Ziółkowska & Mirosław Kobierski**

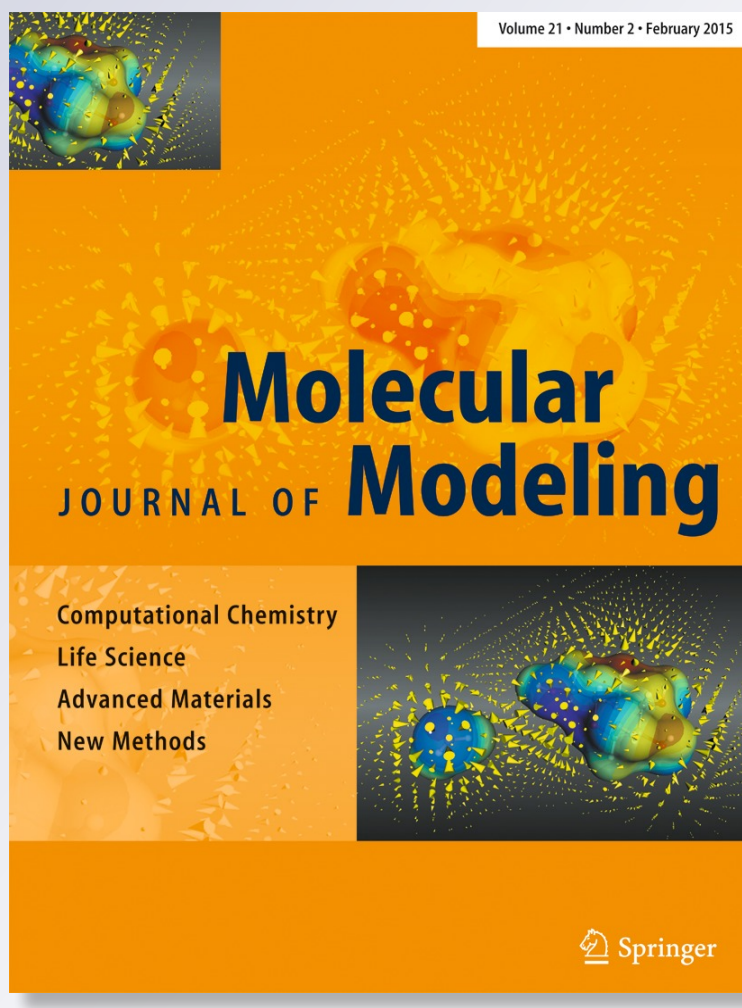

ORIGINAL PAPER

# On the origin of surface imposed anisotropic growth of salicylic and acetylsalicylic acids crystals during droplet evaporation

Maciej Przybyłek · Piotr Cysewski · Maciej Pawelec · Dorota Ziółkowska · Mirosław Kobierski

**Abstract** In this paper droplet evaporative crystallization of salicylic acid (SA) and acetylsalicylic acid (ASA) crystals on different surfaces, such as glass, polyvinyl alcohol (PVA), and paraffin was studied. The obtained crystals were analyzed using powder X-ray diffraction (PXRD) technique. In order to better understand the effect of the surface on evaporative crystallization, crystals deposited on glass were scraped off. Moreover, evaporative crystallization of a large volume of solution was performed. As we found, paraffin which is non-polar surface promotes formation of crystals morphologically similar to those obtained via bulk evaporative crystallization. On the other hand, when crystallization is carried out on the polar surfaces (glass and PVA), there is a significant orientation effect. This phenomenon is manifested by the reduction of the number of peaks in PXRD spectrum recorded for deposited on the surface crystals. Noteworthy, reduction of PXRD signals is not observed for powder samples obtained after scraping crystals off the glass. In order to explain the mechanism of carboxylic crystals growth on the polar surfaces, quantum-chemical computations were performed. It has been found that crystal faces of the strongest orientation effect can be characterized by the highest surface densities of intermolecular interactions energy (IIE). In case of SA and ASA crystals formed on the polar surfaces the most dominant faces are characterized by the highest adhesive and cohesive properties. This suggests that the selection rules of the orientation effect comes directly from surface IIE densities.

## Introduction

Studies on the aromatic carboxylic acids crystal growth and morphology control have considerable pharmaceutical significance and many of these compounds are used as analgesic and anti-inflammatory drugs [1]. Salicylic acid (SA) and its derivative, acetylsalicylic acid (ASA) are probably among the most known representatives of aromatic carboxylic acids being active pharmaceutical ingredients. The curing properties of salicylic acid have been used for a long time. However, due to the side effects it was replaced with aspirin [2]. Nowadays salicylic acid is mainly used in organic synthesis [3], food industry [4, 5], cosmetics and as a component of many cocrystals of pharmaceutical relevance [6–9]. The morphology control belongs to crystal engineering domain, which is very rapidly growing field. Depending on the method of crystallization, one can obtain materials with different properties. This is a very important issue especially in the context of enhancing solubility [10–18], increasing bioavailability

M. Przybyłek (✉) · P. Cysewski · M. Pawelec
Department of Physical Chemistry, Collegium Medicum of Bydgoszcz, Nicolaus Copernicus University in Toruń, Kurpińskiego 5, 85-950 Bydgoszcz, Poland
e-mail: m.przybylek@cm.umk.pl

D. Ziółkowska
University of Technology and Life Sciences in Bydgoszcz, Faculty of Chemical Technology and Engineering, Seminaryjna 3, 85-326 Bydgoszcz, Poland

M. Kobierski
Department of Soil Science and Soil Protection, University of Technology And Life Sciences in Bydgoszcz, Faculty of Agriculture and Biotechnology, Bernardyńska 6, 85-029 Bydgoszcz, Poland [19–24] moisture uptake (hygroscopicity), chemical stability (shelf life), hydrate/solvate formation, crystal morphology, fusion properties, chemical and thermal stability, and mechanical properties of solids [25, 26]. Among many available methods of crystallization, solvent evaporation deserves particular attention. It is worth noting that many papers dealing with evaporative crystallization have appeared recently [27–34]. Although solvent evaporation is one of the simplest methods of crystals production, there have been reported many variations of this technique as for example spin coating [28], spray drying [35], droplet evaporation [36–39], and microwave-accelerated evaporative crystallization [40–42]. The common feature of this approaches is that thin layers or small drops of solutions are evaporated from different surfaces. It is worth mentioning that heterogeneous nucleation on polymer surfaces was found to be a useful method of obtaining protein crystals [43, 44]. Another advantage of this approach was reported by Lee et al. [45] who found that crystallization of celecoxib (a sulfonamide drug) in the presence of polyvinylpyrrolidone leads to formation of crystals with improved drug release behavior. Apart from the practical applications, crystallization on the surface is also interesting from the perspective of the theoretical aspect clarifying nature of crystal growth. It is understandable that, when crystallization is nucleated on the surface, crystals grow is forced in a direction perpendicular to the surface. For this reason there is a significant reduction of peaks number on diffraction spectra measured for crystals deposited on the surface compared to typical spectra of bulk solution crystallization [46–49]. Noteworthy, according to recent studies oriented depositions can also be obtained via nanoparticles suspension drying [50, 51] and with the use of flow-enabled self-assembly method [52].

Studies on the intermolecular interactions in crystals are important from many points of views and are especially valuable in crystal engineering, drug manufacturing, nonlinear optics, high energy density compounds, and supramolecular chemistry [53–59]. Salicylic acid and its derivatives are often used as a model compounds in spectroscopic and theoretical studies on the intermolecular and intramolecular hydrogen bonds [60–63]. It is well known that crystal packing and morphology are closely related to intermolecular interactions stabilizing the crystal lattice. Salicylic acid crystallizes in the monoclinic space group P21/a with a needle-like morphology which indicates strongly anisotropic behavior of crystal faces growth. For rheological reasons, needle-shaped morphology is less convenient for pharmaceutical manufacturing applications than spherical particles [64, 65]. In the case of ASA, two commercially available, prism-shaped, and tetragonal crystals are not suitable for tableting [66]. According to many studies, low-aspect ratio pharmaceutical particles, including aspirin obtained via spherical crystallization exhibit very good tabletability, cohesivity, and flowability [66, 67]. This suggests that anisotropic crystal growth, which leads to high-aspect ratio morphology is undesirable from the viewpoint of practical pills preparation.

Since crystal shape depends on the strength of intermolecular interactions stabilizing particular crystal faces, salicylic acid crystal morphology is crucially related to the intermolecular OH⋯O hydrogen bonds between -OH and -COOH groups as well as between two -COOH groups forming the $C_2^2(8)$ synthon. Aspirin exists in two experimentally observed polymorphic forms sharing the same monoclinic P21/c space group. First crystal structure of aspirin (form I) was reported in 1964 [68]. Interestingly, form II was predicted using computational tools [69] before experimental crystal structure was solved and published in 2005 [70]. It is worth noting that due to the wide conformational landscape of acetylsalicylic acid there are many other theoretical polymorphs which were not identified experimentally [71]. Crystal structure analysis of acetylsalicylic acid polymorphic forms suggests that the main intermolecular interactions are hydrogen bonds involving -COOH groups. There are also weak CH⋯O intermolecular interactions between acetyloxy and phenyl group among other contacts of small contributions. In one of our earlier studies [49] it has been shown that benzoic acid crystal growth on the surface during solvent evaporation can be rationalized in terms of surface density of intermolecular interactions characterizing crystal faces. This approach is based on the additive model which assumes that total energy of the crystal lattice can be calculated by adding all interaction energies of pairs in crystal molecular shell. This simple model was found to be in surprisingly good accord with experimental heat of sublimations. The present paper further explores this idea and deals with surface crystallization of salicylic and acetylsalicylic acids. The main goal of this study is to describe and rationalize crystal growth behavior of these carboxylic acids on different surfaces by means of intermolecular interactions calculations. Particularly it is interesting if cohesive and adhesive properties can be more generally applied for understanding of carboxylic acids crystals growth anisotropy on polar surfaces during droplet evaporation.

## Methods

### Chemicals

Analytical grade of salicylic (SA) and acetylsalicylic (ASA) acids, polyvinyl alcohol (PVA), methanol, and N,N-dimethylformamide (DMF) were purchased from POCH (Gliwice, Poland).

### Glass slides coating

The crystallites of SA and ASA were prepared on microscope slides that were either blank or coated with films of paraffin and PVA. PVA films were obtained by uniformly spreading polymer solution (0.05 g/ml) on the glass plates and drying them under atmospheric pressure at temperature fixed at 43°C. Paraffin coatings were prepared by covering microscope slides with Parafilm "M" (American National Can, Greenwich).

Crystallization and samples preparation procedures

The crystallization of SA and ASA was performed on different surfaces by evaporation of 20 μL droplets of solution under atmospheric pressure at 43 °C. In order to compare diffraction patterns recorded for oriented crystals with those grown on the surface but randomly arranged, crystals deposited on glass were scraped off. This operation was carried out as carefully as possible, taking care that scraping does not significantly affect crystals shape by firmly sliding the razor along the glass surface. The bulk crystallization was performed by evaporation of 30 ml solution at 43 °C.

XRD measurements

Powder X-ray diffraction (PXRD) spectra were recorded using Goniometer PW3050/60 armed with Empyrean XRD tube Cu LFF DK303072. Diffraction data were collected in the range of 2θ between 2° and 40° with 0.001° step width. The patterns were processed in Reflex module of Accelrys Material Studio 7.0 (MS7.0) [72] by sequence of K$\alpha_2$ stripping, background computations and subtraction followed by curve smoothing and normalization.

Calculation details

Cohesive and adhesive properties of SA and ASA were analyzed using intermolecular interactions energy (IIE) surface densities calculated according to the procedure reported previously [49]. In the first step the periodic system geometry optimizations of structures deposited in the Cambridge Structural Database (CSD) [73] were performed using DMol$^3$ [74–76] module implemented in MS7.0 package [72]. This procedure is relied on the molecular geometry relaxation with fixed unit cell parameters to the experimental values. Geometry optimizations were performed using PBE functional [77] and DNP basis set (version 3.5) [78]. It is worth mentioning that DNP is a double numerical basis set which includes d-type polarization function on heavy atoms and p-type polarization function on hydrogen atoms. Such basis set extension along with Grimme [79] dispersion correction is essential for proper hydrogen bonding calculations and stacking interactions. In order to further improve the accuracy of geometry optimization, all-electron core treatment was adopted and integration accuracy, self-consistent field (SCF) tolerance, and orbital quality cutoff were set to fine level. At the next stage the obtained structures were used for preparation of molecular shell (MS) and identification of unique pairs within the crystal. The closest proximity of molecules can be defined by the distance between two molecules, which does not exceed the van der Waals radius augmented by 1 Å [80–84]. The pair intermolecular interaction energies (IIE) were obtained using meta hybrid M06-2X functional [85] along with ET-pVQZ basis set. The basis set superposition error (BSSE) corrections were included. Besides, the nature of intermolecular interactions was studied using Morokuma-Ziegler energy decomposition scheme [86]. According to this approach intermolecular interaction energies can be expressed as a sum of electrostatic ($\Delta E_{EL}^{surf}$), Pauli repulsion ($\Delta E_{TPR}^{surf}$), and orbital interactions contributions ($\Delta E_{OI}^{surf}$) [86]:

$$\Delta E_{IIE}^{surf} = \Delta E_{EL}^{surf} + \Delta E_{TPR}^{surf} + \Delta E_{OI}^{surf}. \qquad (1)$$

The superscript *surf* in above equation denotes that IIE surface densities are calculated by dividing energy values by the surface area of corresponding crystal face. Electrostatic energy contribution is defined by Coulomb forces between unperturbed monomers charge distributions. The Pauli repulsion contribution is associated with the repulsion between occupied orbitals coming from Pauli principle. Orbital interactions contributions stand for HOMO-LUMO charge transfer. This term also includes inner-fragment polarization (mixing of empty and occupied orbitals on one fragment imposed by another one). Intermolecular energies and their decomposition presented in this paper were performed using ADF2013 software [87].

The quantification of cohesive and adhesive properties of crystal faces were estimated according to details previously presented [49]. This three stages procedure started from identification of the most important faces based of given crystal by the analysis of PXRD spectra generated in Reflex module of MS7.0. The values of Miller indices of the most dominant faces were used for cleaving procedure leading to corresponding surfaces. All possible positions of cleaving were considered according to functionalities of MS7.0. Finally the intermolecular interaction profiles were computed and used for characteristics of cohesive and adhesive properties. This was done by generation of 3x3x1 cells and using them as building blocks for infinite surfaces construction. Practically, the surface exposed cells (SEC) of 9x9 size were sufficiently large for mimicking infinitely extended surfaces and elimination edges artifacts. Cohesiveness of crystal faces is defined as intra-layer IIE densities of interactions between carboxylic acids molecules found within SEC along 2D directions. Adhesive properties of crystal faces are associated with inter-layer interactions of carboxylic acids molecules in SEC with its images along in-depth direction.

## Results and discussion

Crystallization conditions selection

It is well known that crystal size and shape can be affected by crystallization conditions. Among important factors influencing crystallization there are both types of solvent and solution concentration. As we found concentration of SA and ASA used in droplet evaporative crystallization does not significantly affect relative intensities of signals on the diffraction pattern recorded for that deposited on the surface crystallites due to the fact that utilized crystallization conditions lead to a very rapid process at highly non-equilibrated crystals formation from saturated solution. It is worth mentioning that too low concentration leads to the formation of crystallite layers which are not sufficiently thick for recording a good quality PXRD spectra. Noteworthy, it is possible to reduce the noise in diffraction spectra through multiple dropping and evaporating the solution on the surface. It was found that this procedure does not affect crystal faces orientation seen through PXRD patterns. This is of particular importance for poorly soluble substances. However, in this study concentration of 0.724 M was used which allows to obtain a good quality PXRD spectra of SA and ASA crystallites after a single droplet evaporation. Presented herein crystallization conditions provide morphologically uniform crystal layers which was additionally established by optical microscope observations.

In order to evaluate the effect of the type of solvent on the crystallites morphology two different solvents were used, namely methanol which is volatile (boiling point, b.p.= 65 °C) and protic solvent and DMF which is aprotic and of a low volatility (b.p.=153 °C). In Fig. 1 exemplary powder diffraction patterns recorded for SA crystallites deposited on glass were summarized. As one can see, the type of solvent used in the crystallization does not significantly affect PXRD spectra recorded for obtained crystallite layers and hence crystallization from methanolic solutions was selected for further studies. This observation is interesting, because it shows that such solvent features like volatility and proton-donating/ accepting ability does not affect orientation of SA and ASA crystal faces deposited on the surface.

Evaporative crystallization

The initial stage of this study was devoted to measurements of influence of the surface type on the morphology of SA and ASA crystals. For this purpose the evaporation of small droplets of methanolic solution on polar and easily wettable surfaces (glass and PVA) and on non-polar surface, paraffin were performed. The representative microscope images of obtained crystallites were collected in Fig. 2. It is immediately evident from these pictures that the morphology of fully developed crystals obtained through evaporation of a large amount of solution in bulk condition are completely different from those growing on the surfaces. Also crystals obtained through droplet evaporative crystallization of SA and ASA on highly wettable surfaces differ from those obtained on non-polar film. It is interesting to note that polar surfaces promote formation of similar flat longitudinal crystals arranged close to each other on the surface. Such qualitative morphology characteristics was augmented by quantitative analysis via X-ray powder diffraction measurements. The PXRD spectra presented in Figs. 3 and 4 correspond to reflections from the surface covered by crystals. Since the pure paraffin films exhibit some degree of crystalline properties, the PXRD spectra characterizing this coating were also presented. Apart from the surface crystallization, there were also recorded spectra for crystals obtained via classical bulk crystallization as well as ones coming from carefully scrapping off the crystals from the surfaces. In Figs. 3 and 4 there are also provided idealized spectra. These spectra were generated in Reflex module based on the contents of CIF files deposited in Cambridge Structural Database (CSD).

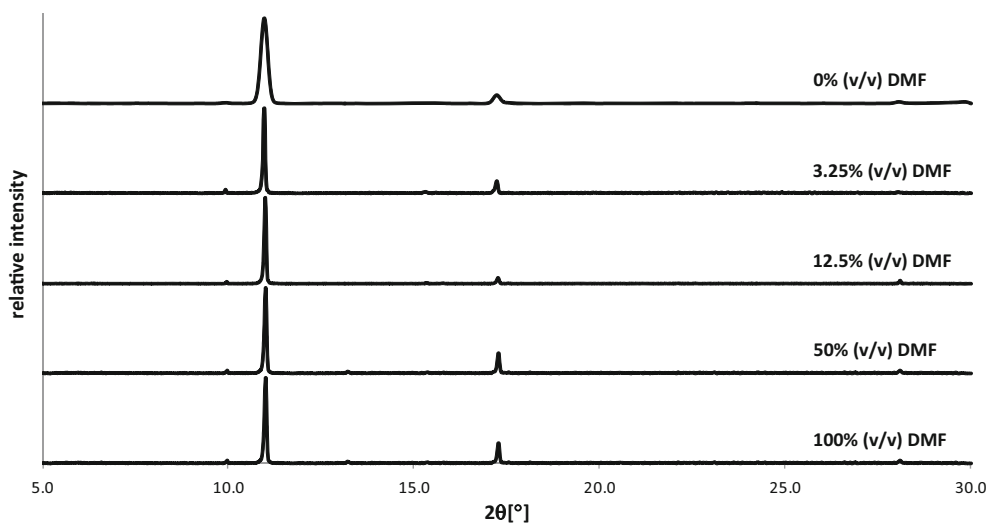

Fig. 1 PXRD spectra recorded for that deposited on the glass crystallites obtained through droplet evaporation of SA dissolved in methanol, and in several methanol/DMF mixtures

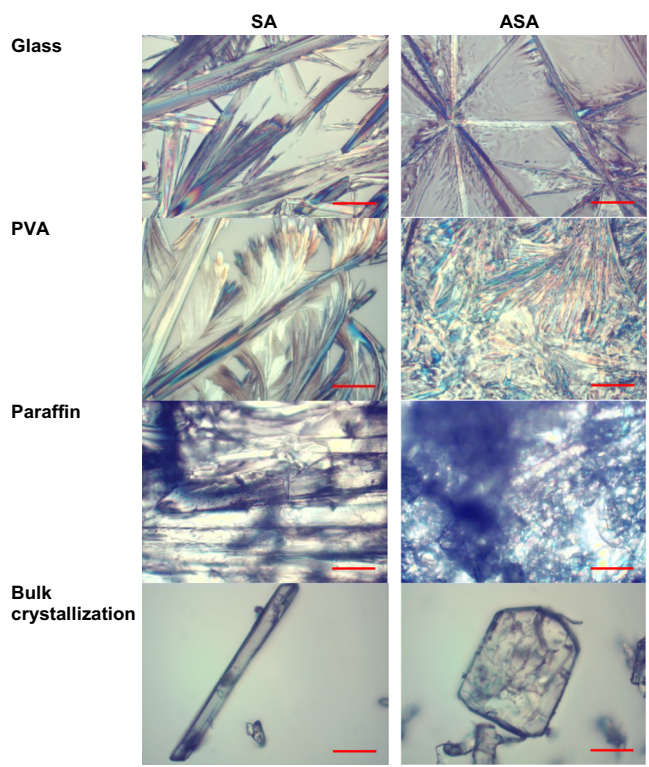

**Fig. 2** The representative microscope images of salicylic acid (SA) and acetylsalicylic acid (ASA) crystals formed on different surfaces. Red scale bar corresponds to 50 μm

As can be inferred from Fig. 3, the bulk crystallization of SA leads to full development of crystal morphology and in applied crystallization conditions several faces typical for this solid were observed. The three most significant reflexes correspond to (210), (110), and (121) faces. There is direct correspondence of obtained signals with idealized ones. The most interesting observation coming from performed experiments is the significant reduction of the number of measured signals after surface crystallization. This is especially spectacular in the case of on-glass crystallization, for which practically one signal was detected at 2θ=11.0°. This suggests that at experimental conditions the morphology of salicylic acid crystallites is overwhelmingly dominated by (110) face. This is also typical for other polar surfaces as PVA, although in this case also (210) face is detectable. It is interesting to inspect spectra obtained after carefully scrapping crystals off the glass surface. In such case the PXRD patterns are quite similar to ones characterizing products of bulk crystallization. This suggests that the observed extremely strong influence of polar surfaces on the crystal growth during droplet evaporation is associated mainly with directional effect. The crystals are orientated and tightly packed on the surface. Since the film of crystallized salicylic acid is thin and nearly homogenous the detected reflections of X-ray diffraction are strong and detectable only for (110) Miller plane. On the other hand, this particular surface is exposed outward and physicochemical properties of such crystallite are mainly defined by this particular face.

The PXRD spectra of the second compound studied here, aspirin were presented in Fig. 4. Although ASA crystals exist in two polymorphic forms (reported in CSD as ACSALA (polymorph I) and ACSALA17 (polymorph II) in experimental conditions utilized in this study only the former form is obtained. Indeed diffraction patterns of polymorph I and II differ so significantly that it is hardly possible to misinterpret them. Thus, in the case of bulk crystallization from methanol solution aspirin develops crystals with morphology typical for polymorph I. The constraining of multidimensional crystallite growth on the polar surfaces leads to significant reduction of signals and in the case of glass surface practically only two faces are detectable. The most intense signals came from (100) and (002)/(200) Miller planes. The latter two are indistinguishable due to high overlap. It is worth mentioning that atomic force microscopy (AFM) and photoelectron spectroscopy measurements showed that ASA crystal faces reveal different hydrophilicity depending on the type of functional groups terminating the surface [88, 89]. The (100) crystal face which is dominant on polar surfaces is significantly much more hydrophilic than (002), (011), and (110) faces [90]. This observation and presented herein method of ASA crystallization on polar surfaces can be useful in the context of thin layers production with enhanced dissolution properties. The comparison of pattern obtained on glass surface with the one obtained after scrapping crystals of the surface leads to similar concussion as drawn from salicylic acid crystallization. Hence, the source of reduction of diffraction signals lies in a very strong directional effect of polar surfaces. The PXRD spectrum measured for crystals deposited on paraffin are more similar to spectra recorded for crystals obtained via bulk evaporative crystallization. Due to the low wettability of this surface ASA crystals faces growth proceeds without directional effect, contrary to crystallization on the polar surfaces.

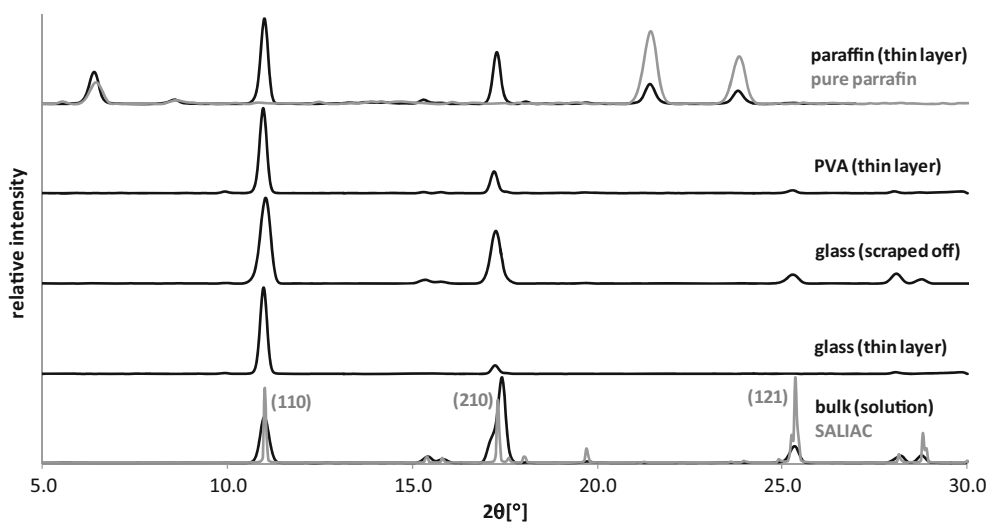

**Fig. 3** PXRD spectra recorded for SA crystals deposited on glass, PVA, and paraffin

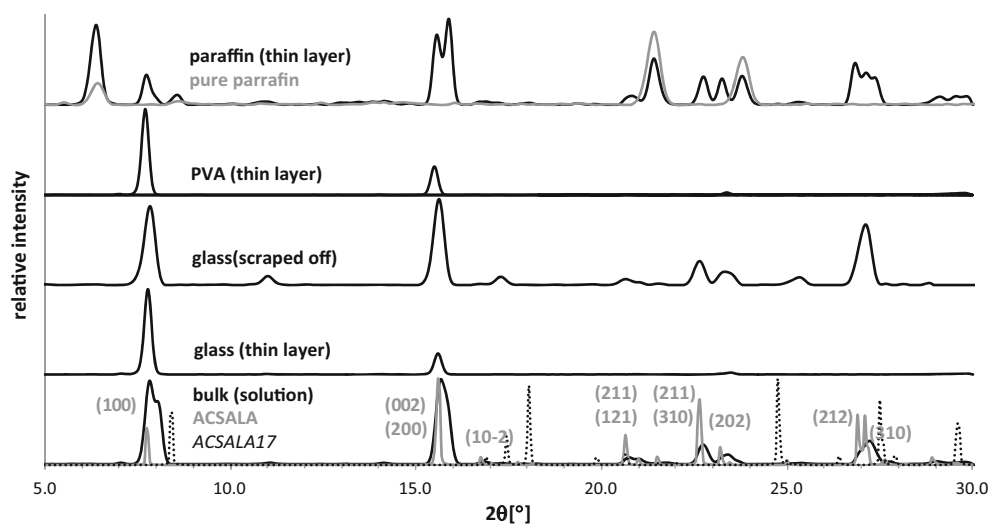

**Fig. 4** PXRD spectra of ASA crystal layers formed on glass, PVA, and paraffin

Energetic stabilization of crystal faces

The above-presented observations of highly anisotropic behavior of crystal growth on polar surfaces, which is particularly emphasized by glass surfaces, raises a thought-provoking question about the origin of the orientation effect. It is interesting to know what the selection rules are for promoting crystals growth toward particular directions in given conditions. For this purpose the intermolecular interactions were analyzed in detail for each of the potentially important faces. Since the arrangements of molecules are specific to crystal surfaces, the interactions of cohesive and adhesive nature are also expected to be face related. In the previous paper [49] parallel analysis was performed for benzoic acid crystals developed in analogical conditions. Similar methodology adapted here enables for the analysis of the role of intermolecular interactions stabilizing SA and ASA crystals. Since both systems are characterized by one molecule per asymmetric unit (Z'=1) all molecules are supposed to be structurally and energetically identical in either crystal. There are ten deposits of SA measured in the temperature span from 90 K up to ambient one. For the purpose of this study the crystal structure of SA deposited in CSD under code SALIAC was the starting point for partial optimization. In SA crystal each molecule is surrounded by 15 neighbors belonging to the molecular shell but only nine intermolecular contacts are distinct. They univocally define all energetic patterns within both bulk crystal and morphology related faces. The values of selected contributions to intermolecular interactions of these contacts are provided in Table 1. Similar characteristics were also provided for aspirin. The CSD comprises 22 records of ASA crystals in two polymorphic forms. The supplied deposits correspond to those measured in the range of temperature from 20 K up to 300 K. As one can infer from deposited data for both aromatic acids there is almost linear temperature dependence of cell volume rising with increase of temperatures. The structure of the most stable polymorphic form at ambient conditions, ACSALA, was used for further analysis. The nine unique energetic patterns are characterized in Table 1. Both aromatic acid crystals form $C_2^2(8)$ synthon, which IIE dominates among all other. Interestingly, synthon interaction is slightly stronger in the case of ASA compared to SA pairs. Data provided in Table 1 can be directly used for computation of lattice energy, $E_{latt} \approx \Delta E_{MS} = 0.5 \cdot \Sigma n_{ij} \cdot \varepsilon_{IIE}$ (ij) (0.5 factor is used in order to avoid double counting of intermolecular interactions in crystal). One can evaluate the accuracy of additive model by comparison to experimental values of sublimation enthalpies. Since the sublimation enthalpy, $\Delta H_{sub}$ (T), of a crystal is a direct measure of the lattice energy, these data are very often used for theoretical models verification [91, 92] by the following simplified formula: $\Delta H_{sub}(T) = -E_{latt} - 2RT$, where T is the temperature at which the sublimation enthalpy is measured and R stands for the gas constant. The experimental values of $\Delta H_{sub}$ (T) are available [92]. The average value of SA crystal lattice energy is equal to −23.99 kcal mol$^{-1}$. The thermochemical data characterizing sublimation of ASA are also available [93] and equals −34.17 kcal mol$^{-1}$. In the case of applied additive model the predicted stabilization energies are −24.06 kcal mol$^{-1}$ and −33.23 kcal mol$^{-1}$ for SA and aspirin, respectively. Although additive model under-stabilize aspirin crystals the agreement with experimental data is quite satisfactory. Thus, the additive model was used for characterizing surface densities of energies according to the approach described in methodology section. Based on PXRD spectra generated using CIF files the relative intensities of signals (I/Imax) were obtained. For further analysis those faces were considered which were no less intense than 5 %, as compared to the highest peak. This led to inclusion of 12 faces for SA and 26 Miller planes for ASA, respectively. For all selected faces the intermolecular interaction profiles were computed for characteristics of adhesive and cohesive properties. This analysis enabled not only for quantification of the intermolecular interaction densities but also detailed analysis of contributing factors coming from energy decomposition analysis. The results of such analysis are provided in Fig. 5.

**Table 1** The unique pair interactions defining intermolecular interactions stabilizing SA and ASA crystals. All energies (in kcal mol$^{-1}$) were computed at M06-2X/ET-pVQZ level using ADF software [87]

| SALIAC | | | | | |
|---|---|---|---|---|---|
| -x,-y,-z | −21.82 | 41.36 | −36.50 | −26.68 | 4.86 |
| -x,-y,-1-z | −4.67 | −0.73 | −2.41 | −1.53 | −3.14 |
| x,y,-1+z | −3.68 | −0.47 | −1.61 | −1.60 | −2.09 |
| −1/2+x,1/2-y,z | −2.68 | 0.27 | −2.01 | −0.94 | −1.74 |
| 1/2-x,-1/2+y,-2-z | −1.38 | 0.62 | −1.44 | −0.56 | −0.81 |
| 1/2-x,-1/2+y,-1-z | −1.38 | 0.02 | −0.76 | −0.63 | −0.75 |
| -x,1-y,-1-z | −1.12 | 0.00 | −0.62 | −0.50 | −0.62 |
| −1/2+x,1/2-y,1+z | −0.99 | 0.79 | −1.07 | −0.70 | −0.28 |
| -x,1-y,-2-z | −0.29 | 0.38 | −0.04 | −0.63 | 0.34 |
| ACSALA | | | | | |
| -x,-y,-z | −24.31 | 44.31 | −39.84 | −28.78 | 4.47 |
| x,-1+y,z | −4.46 | 0.37 | −3.27 | −1.56 | −2.90 |
| 1-x,1-y,-z | −4.36 | −0.88 | −1.60 | −1.88 | −2.49 |
| -x,1-y,-z | −4.33 | −1.08 | −1.41 | −1.83 | −2.50 |
| 1-x,-y,-z | −3.81 | 1.90 | −4.36 | −1.34 | −2.47 |
| x,1/2-y,-1/2+z | −2.91 | 0.58 | −1.93 | −1.56 | −1.35 |
| -x,-1/2+y,1/2-z | −1.95 | 1.00 | −1.82 | −1.13 | −0.82 |
| x,1.5-y,-1/2+z | −1.62 | −0.27 | −0.65 | −0.70 | −0.92 |
| 1-x,-1/2+y,-1/2-z | 0.03 | −0.08 | 0.34 | −0.23 | 0.26 |

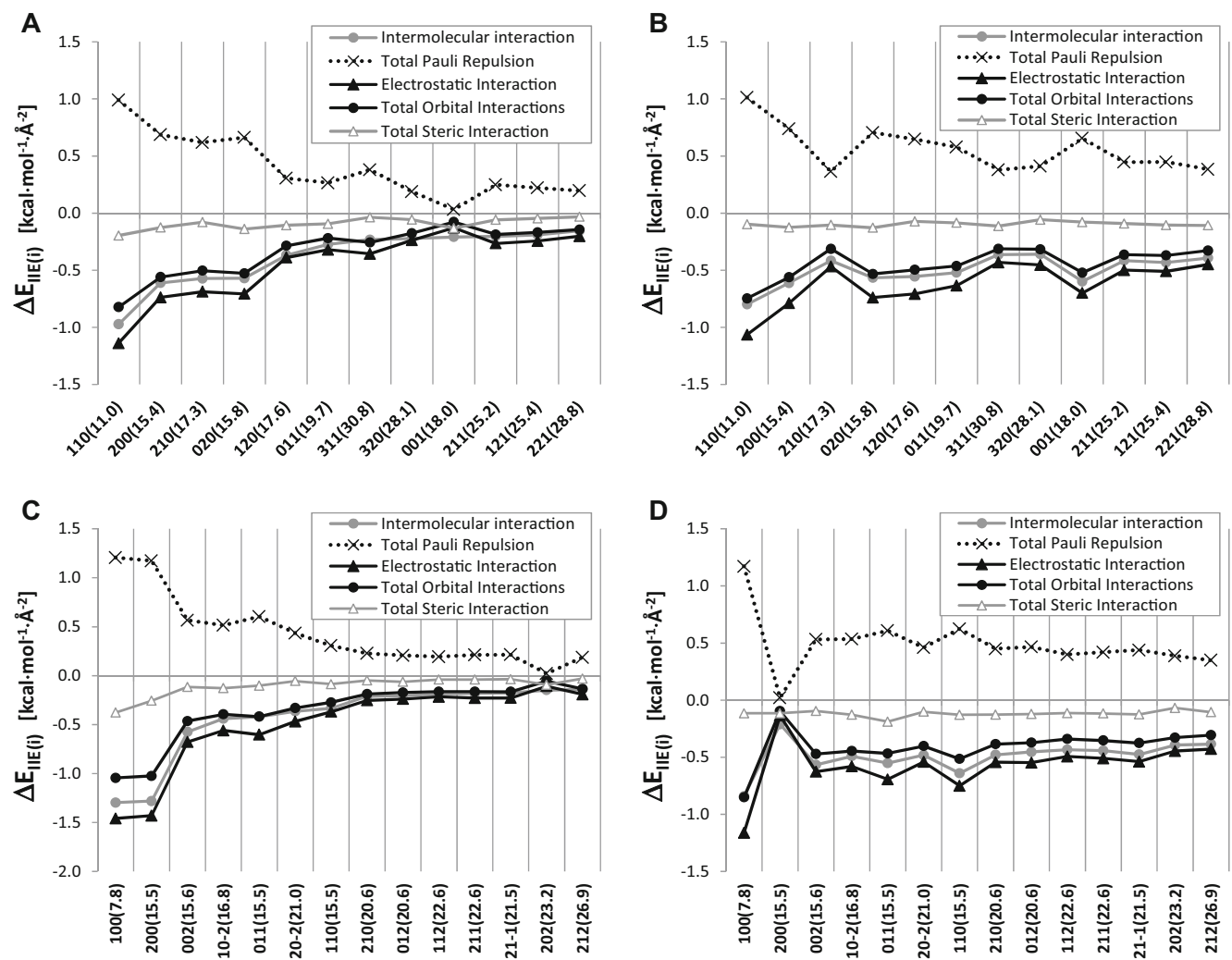

**Fig. 5** Intra-layer (**a**) and inter-layer (**b**) intermolecular interaction energies (IIE) densities of SA crystal faces and intra-layer (**c**) and inter-layer (**d**) IIE densities of ASA crystal faces. Miller indices corresponding to each face and 2θ values (in parenthesis) are provided in description of abscissa

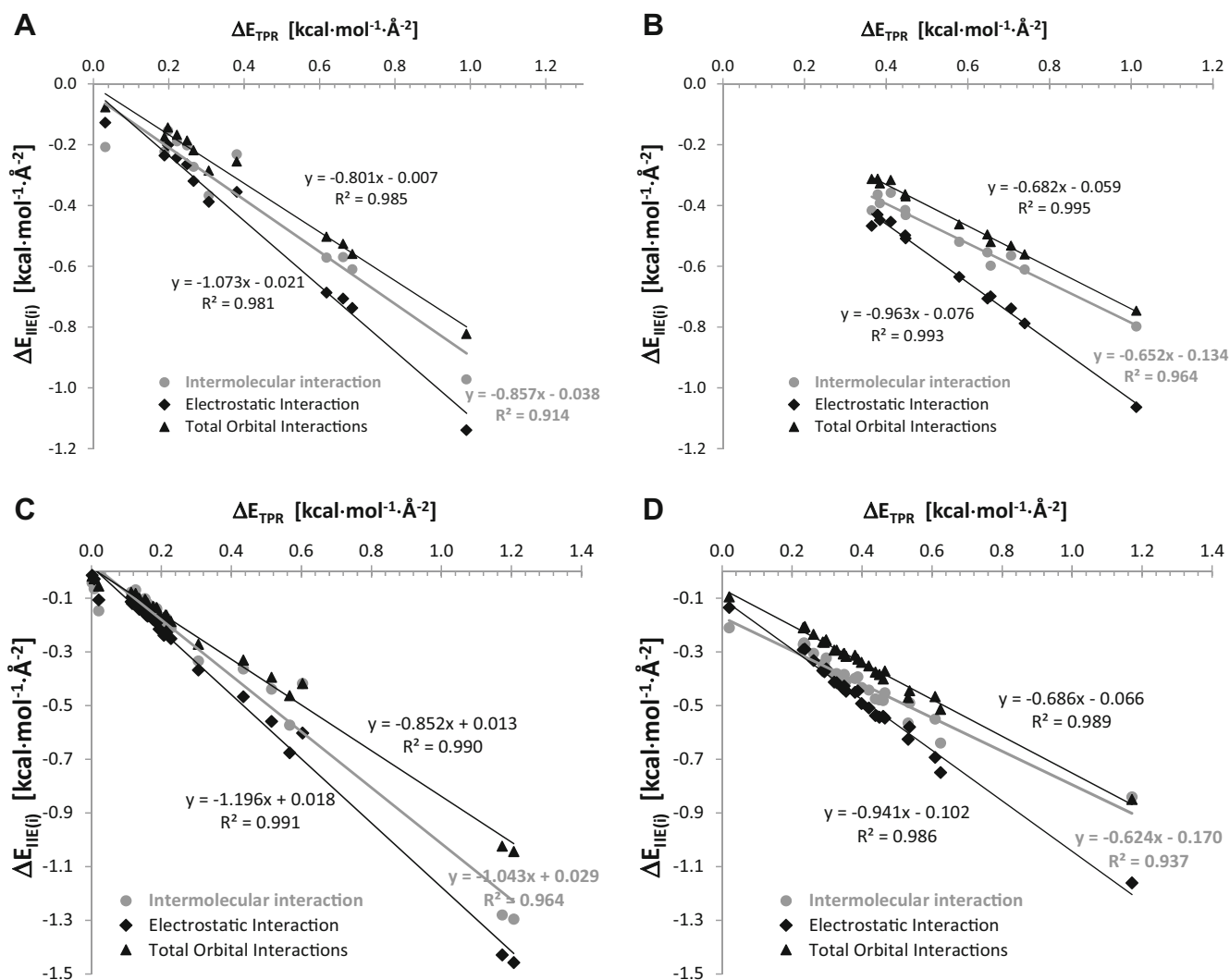

**Fig. 6** Linear relationships between repulsive and attractive contributions to (a) intra-layer and (b) inter-layer IIE densities of SA and to (c) intra-layer and (d) inter-layer IIE densities of ASA

As was documented in Fig. 5, the experimentally observed growth of carboxylic acid faces on polar surfaces is closely associated with anisotropy of cohesive and adhesive properties of crystal. Interestingly this experimental observation is in good accords with surface densities of intermolecular interactions. Indeed as it is documented in Figs. 5a and c the faces found in crystallites of SA and ASA on glass films are those with highest cohesive character. This is also associated with the smallest surface unit cell. Thus, the selection rule for directing of crystallization on polar surfaces is quite simple and very intuitive. The reduction of the total surface during crystallization is associated with exposure of SEC of lowest area, which in turn have the strongest densities of intramolecular interactions. As has been found, in the case of ASA, (100) plane is the most energetically stable in terms of both cohesive and adhesive properties (Figs. 5c and d). Apparently, this face corresponds to the most dominant peak on PXRD spectra recorded for ASA crystals deposited on the polar surfaces. In the case of SA the (110) face exhibits similarly highest cohesive and adhesive properties (Figs. 5a and b) and also is observed experimentally. The conclusion that IIE densities control orientation effect on the polar surfaces was also documented for benzoic acid [49]. One can anticipate that the generalization of these observation is valid and for other crystals it is expected that those surfaces having of both the strongest cohesive and adhesive properties are ones appearing predominantly on the glass surfaces during droplet evaporation crystallization.

The inspection of energetic contributions to IIE presented in Fig. 5 suggests that the electrostatic contribution is generally the most important in the context of crystal faces stabilization. Since the cohesive and adhesive properties of carboxylic acids crystals are predominantly affected by hydrogen bonds found in $C_2^2(8)$ synthon the hydrogen bonds nature is responsible for observed face related IIE. On the other hand, almost symmetrical arrangement of total Pauli repulsion plot with respect to the other plots suggests that repulsive and attractive contributions are not independent from each other. Indeed, linear correlation been found between Pauli repulsion surface density energy and attractive interactions contributions with very high correlation coefficient. The observed trend was documented in Fig. 6. As one can see from the slope values analysis, total Pauli repulsion increase is generally more sensitive to the increase of attraction coming from total orbital interactions contribution than originating from electrostatic contribution.

## Conclusions

The salicylic acid and aspirin are very important compounds both from pharmaceutical industry, cosmetics, and organic synthesis. They are used as curing agents in a number of diseases as anti-inflammatory agents, easing aches and pains, key ingredients in many skin-care products [94]. Thus, controlled crystallization of these chemicals is important both from practical and theoretical points of view. In general any crystal growth is of anisotropic nature since crystal faces have different growth rates. In this study such anisotropy of SA and ASA crystallization was intentionally enhanced by restricted crystallization via droplet evaporation on different surfaces. In the case of polar surfaces, including glass and PVA, the number of peaks registered by XRPD measurements was dramatically reduced. This has been attributed to the fact that crystals were force to grow rapidly at elevated temperature in the direction perpendicular to the surface. Consequently only certain faces were able to be exposed outward. The other faces are not directly detectable in the powder pattern due to strong orientation effect associated with directional arrangement of crystals on the surface. In order to evaluate the influence of a non-polar surface on SA and ASA crystals growth behavior, droplet evaporative crystallization on paraffin layer was performed. In case of this coating, bulk-like crystal morphology of both SA and ASA was observed. This is probably caused by the fact that solution layer on low wettable surface is thicker and crystals can grow in all directions. Probably, because of the lower surface area of droplet on paraffin in comparison to glass and PVA surfaces, the orientation effect is less pronounced on paraffin. Anisotropic crystal growth is closely associated with high-aspect ratio morphology which is not suitable for drug manufacturing. For this reason SA and ASA crystal growth on non-polar surface is probably more suitable for pharmaceutical manufacturing applications than crystallization on polar surface. On the other hand, in the case of ASA crystallization on glass and PVA surfaces, hydrophilic (100) crystal face [90] is exposed. This is interesting in the context of promoting formation of crystallite layers with enhanced solubility.

In order to explain the origin of observed phenomenon of carboxylic crystals growth on the polar surfaces, quantum-chemically derived intermolecular interaction were used. It has been found that surface densities of IIE can provide an understandable explanation of unique properties of crystal faces. According to this approach two types of IIE surface densities can be distinguished, namely cohesiveness and adhesiveness. The first of these features characterizes intra-layer interactions in crystal face, the latter inter-layer interactions. It has been found that those crystal faces which correspond to the most intense diffraction peaks exhibit the highest cohesiveness and adhesiveness. The analysis presented here does not include specific interactions between polar surfaces and carboxylic acids molecules. Nevertheless, good results in predicting which crystal face is exposed on the polar surfaces suggest that such a simplified approach provides reliable explanation of observed phenomenon.

In the case of analyzed crystals the electrostatic interactions are the most important factors contributing to both cohesive and adhesive properties of studied carboxylic acids crystals. The noticed very good correlations between total Pauli repulsive densities and other contributions can be regarded as a general feature of applied analysis. The above conclusions are generally consistent with previously reported results for non-substituted benzoic acid [49]. This shows that, anisotropic properties of aromatic carboxylic acids crystals can be described in terms of IIE densities and the intermolecular interactions additive model.

**Acknowledgments** This research was supported in part by PL-Grid Infrastructure. The allocation of computational facilities of Academic Computer Centre "Cyfronet" AGH / Krakow / POLAND is also acknowledged. Valuable technical assistance of Tomasz Miernik is acknowledged.